# A bioimpedance-based cardiovascular measurement system

Roman Kusche[1][0000-0003-2925-7638], Sebastian Hauschild[1], and Martin Ryschka[1]

[1] Laboratory of Medical Electronics, Luebeck University of Applied Sciences, 23562 Luebeck, Germany
`roman.kusche@fh-luebeck.de; martin.ryschka@fh-luebeck.de`

**Abstract.** Bioimpedance measurement is a biomedical technique to determine the electrical behavior of living tissue. It is well known for estimating the body composition or for the electrical impedance tomography. Additionally to these major research topics, there are applications with completely different system requirements for the signal acquisition. These applications are for example respiration monitoring or heart rate measurements. In these cases, very high resolution bioimpedance measurements with high sample rates are necessary. Additionally, simultaneous multi-channel measurements are desirable.

This work is about the hardware and software development of a 4-channel bio-impedance measurement system, whereat all channels are galvanically decoupled from each other. It is capable of measuring 1000 impedance magnitudes per second and per channel. Depending on the chosen measurement configuration, impedance changes down to mΩ ranges are feasible to be detected. To enable the usage in a variety of different research applications, further biosignals like photoplethysmography, electrocardiography or heart sounds can be acquired simultaneously. For electrical safety purposes, an implemented galvanically isolated USB interface transmits the data to a host PC. The impedance measurements can be analyzed in real-time with a graphical user interface. Additionally, the measurement configuration can easily be changed via this GUI.

To demonstrate the system's usability, exemplary measurements from human subjects are presented.

**Keywords:** Bioimpedance, Multi-Channel, Impedance Cardiography, Heart Sounds, Pre-Ejection-Period.

## 1 Introduction

Measuring the electrical bioimpedance has been proven to be a useful method to determine the electrical behavior of living tissue [1]. It is mostly used to estimate the composition of the human body. Further applications are monitoring of the respiration and the cardiac output as well as the detection of arterial pulse waves and muscle contractions [2,3,4,5,6]. For these purposes, several specific bioimpedance measurement systems have been developed and published in the past [3,5,7].

It is conceivable that the additional acquisition of further biosignals like the electrocardiogram (ECG), the photoplethysmogram (PPG) or phonocardiogram (PCG) enables a variety of biomedical measurement setups. A challenge of implementing a multi-channel bioimpedance setup is the electrical decoupling of the different measurement channels from each other and from the ECG, as well as the synchronized data acquisition.

This work proposes the design of a measurement system capable of acquiring up to four bioimpedances simultaneously using galvanically decoupled current sources. Additionally, the synchronized measurement of ECG, PPG, PCG and pressure signals is possible. The microcontroller based device acquires 1000 bioimpedance magnitudes per second and per channel as well as the other mentioned signals with a sampling rate of 1000 samples per second (SPS). The digitized data is transmitted, via an isolated USB port to comply with electrical safety requirements according to IEC 60601-1.

After explaining the respective biosignals, the development of the measurement system is described. Finally, an exemplary measurement setup is proposed to analyze the timing relationship between the ECG, the PCG and the actual blood ejection of the heart.

## 2     Materials and Methods

### 2.1     Measurement Methods

*Bioimpedance Measurement*
To determine the electrical impedance of tissue, a small known AC current ($I_{Meas.}$) in mA ranges is applied via electrodes ($Z_{ESI1}$, $Z_{ESI2}$) to the bioimpedance ($Z_{Bio}$) under test, as shown in Fig. 1. Typical excitation frequencies are in the range of tens to hundreds of kHz [1]. The occurring voltage drop is measured utilizing two additional electrodes ($Z_{ESI3}$, $Z_{ESI4}$).

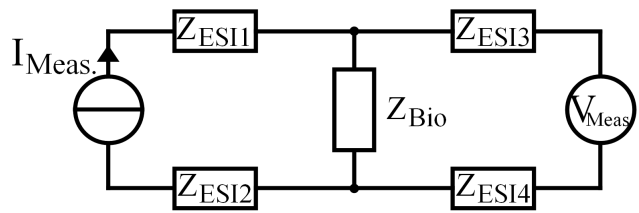

**Fig. 1.** Equivalent circuit of a bioimpedance measurement.

Since the inner impedance of the used voltage measurement circuit ($V_{Meas.}$) is supposed to be extremely high, the voltage drop over $Z_{ESI3}$ and $Z_{ESI4}$ can be neglected and the measured voltage together with the known current $I_{Meas.}$ can be used to calculate the bioimpedance under test.

*Electrocardiography*

The electrocardiography is the measurement of the electrical heart activities. Its morphology can be analyzed to gather information about the heart. Additionally, it can be used as time reference in several applications such as pulse wave velocity measurements [8]. The frequency range of a typical ECG signal is from 0.5 Hz to 100 Hz and the signal's amplitude is between 100 µV and 10 mV [9].

*Phonocardiography*

Phonocardiography is the measurement of the heart sounds, which occur due to the mechanical activity of the heart. Especially the characteristic sounds $S_1$ and $S_2$, caused by the closing of the mitral and aortic valve, are of interest [8].

*Photoplethysmography*

The photoplethysmography is an optical method to acquire the arterial pulse wave. It is based on the measurement of the optical reflection/absorption of light in tissue over time. By performing this measurement using two light sources of different wave lengths, the blood oxygenation can be determined [8].

*In-Ear Pressure*

In-Ear pressure measurements have been proven to be a useful method for detecting the arterial pulse wave [10]. By sealing the ear canal and measuring the occurring pressure changes inside, the pulse wave can be acquired [11].

### 2.2  System Design

To enable the simultaneous measurement of all proposed methods, a specific microcontroller based measurement system has been developed. As it can be seen from the block diagram, illustrated in Fig. 2, the system consists of four voltage-controlled current source (VCCS) modules [12] in combination with a data acquisition (DAQ) module.

The ECG block of the DAQ module is realized by an instrumentational amplifier (INA, INA126 from Texas Instruments) in combination with a driven right leg circuit to attenuate the occurring common mode interferences. To reduce the signal drifting, the output signal of the INA is integrated and fed back to its reference input. In order to remove the high frequency voltages, caused by simultaneous bioimpedance measurements, the ECG signal is filtered by a passive 1st order low pass filter with a cut-off-frequency of $f_c$=600 Hz.

To acquire a phonocardiogram via an electret microphone, a microphone amplifier block (MicAmp) is implemented. For extracting just the useful frequency components of the signal, a 1st order high pass filter ($f_c$=16 Hz) and a 1st order low pass filter ($f_c$=3.6 kHz) are implemented. Additionally, the signal is amplified with a gain of G=20.

The circuitry of the PPG block is split into two parts. The light transmitting part is realized by an infrared LED (VSMB3940X01 from Vishay), whose current is limited to 17 mA. To convert the received light into a voltage signal, a photo diode (VBPW34S from Vishay) in combination with a transimpedance amplifier circuit is provided.

As in-ear sensor, a miniature pressure sensor (HCEM010DBE8P3 from Sensortechnics) is used. It converts the measured differential pressure in a range of ±10 mbar into a corresponding DC voltage between 0.25 V and 2.25 V.

To measure up to four bioimpedances, the data acquisition module controls the four external current source modules. In order to avoid coupling effects, these modules are galvanically isolated from each other, regarding their power supplies as well as the data transmission paths [12]. The configured current sources apply AC currents in the range from 0.12 to 1.5 mA with frequencies between 12 and 250 kHz into the tissue under observation. The occurring differential voltage drops over the bioimpedances are amplified by the differential programmable gain amplifiers (AD8250 from Analog Devices) of the data acquisition board. Since the useful information of these AC signals is their envelope, caused by the bioimpedance magnitude, analog amplitude demodulation circuits are implemented. These circuits are based on analog rectifiers in combination with 6$^{th}$ order active low pass filters ($f_c$=1 kHz).

The resulting eight analog signals are simultaneously digitized by an 8 channels analog-to-digital converter (ADC, ADS131E08 from Texas Instruments) and transmitted to the implemented 32 bits microcontroller (ATSAM4S16C from Microchip). This controller communicates via a galvanically isolated (ISO7721 from Texas Instruments) USB-UART interface with a host PC. The usage of this isolator as well as the usage of a medical power supply is intended to comply with the standard for medical electrical equipment (IEC 60601-1).

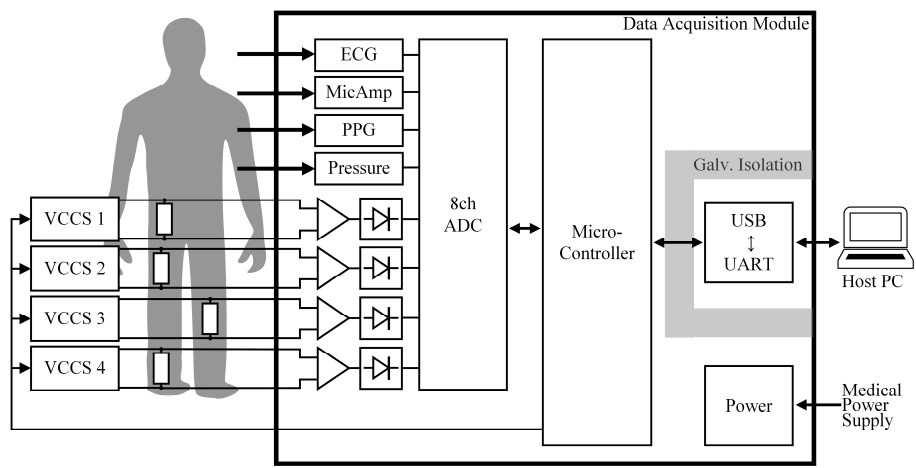

**Fig. 2.** Block diagram of the developed cardiovascular measurement system. The system consists of four galvanically decoupled current source modules (VCCS) and a data acquisition module.

In Fig. 3 a photograph of the developed measurement system, consisting of the data acquisition module and the four decoupled current sources, is depicted. The 4-layers printed circuit board (PCB) of the data acquisition module has dimensions of 161 x 97 mm² and contains more than 600 electronic components.

A graphical user interface, written in C# language, enables the configuration of the decoupled current sources and the programmable gain amplifiers in real-time. Additionally, all acquired signals can be plotted and exported to Matlab (from Math-Works).

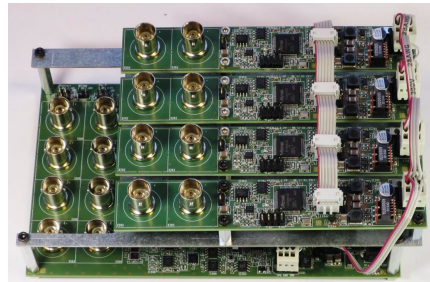

**Fig. 3.** Photograph of the cardiovascular measurement system. The current source modules are mounted on the data acquisition module with a metal sheet for shielding in between.

## 3  Results

To demonstrate the system's usability, an exemplary measurement setup as shown in Fig. 4 (a) has been used for analyzing the timing behavior of the blood ejection of the heart. Therefore, the measurements of an ECG, the heart sounds and the bioimpedance have been performed on a young male subject simultaneously. To acquire the ECG and the bioimpedance signals, Ag/AgCl electrodes were used. In Fig. 4 (b) the acquired signals are plotted for a time duration of 3.5 s.

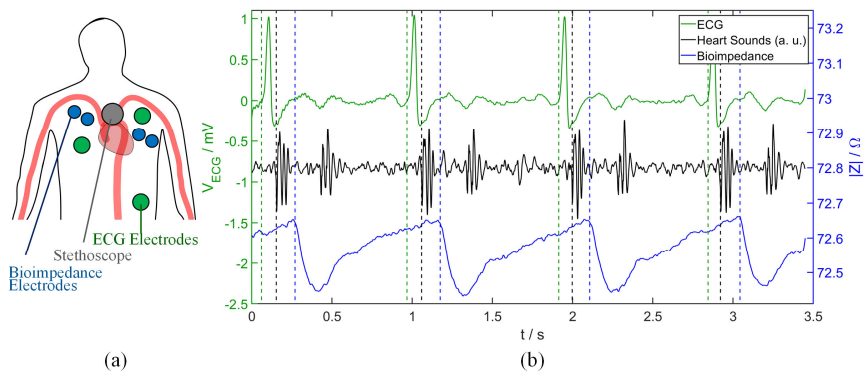

**Fig. 4.** Measurement setup (a) to acquire the ECG, the heart sounds as well as the bioimpedance from the chest. The resulting filtered signals are plotted (b) for a duration of 3.5 s.

To remove the 50 Hz noise, caused by the mains, a digital notch filter was applied to the ECG signal (green). To attenuate high frequency noise, the signal has also been filtered by a 1$^{st}$ order low pass ($f_c$=20 Hz). The same low pass is also applied to the bioimpedance signal (blue). The heart sounds signal from the stethoscope (black) has not been digitally filtered, but normalized for better display. To remove the effect of the filters' delays, only zero-phase filters have been used.

The vertical dashed lines indicate the corresponding cardiovascular events. In the ECG plot this represents the beginning of electrical stimulation. Afterwards, the first heart sound occurs, which is caused by the closing of the mitral valve. Finally, the bioimpedance decreases abruptly. This time point represents the sudden increase of blood volume in the aortic arch, caused by the opening of the aortic valve.

The acquired signals can be used to determine cardiovascular parameters like heart rate, heart rate variability and pre-ejection period.

## 4    Summary and Outlook

In this work, the development of a cardiovascular measurement system has been presented. The system is capable of acquiring up to four bioimpedances, ECG, heart sounds, PPG and In-Ear pressure signals, simultaneously.

A measurement setup to analyze the timing behavior of the heart's blood ejection has been proposed and an exemplary measurement was shown. Further analysis of the resulting ECG, heart sound and bioimpedance signals could be used to measure the pre-ejection period, the heart rate or the ejection duration of the heart.

In the future, algorithms have to be developed to extract the information from the signals automatically. Furthermore, the PPG and the In-Ear pressure signals could be used to measure a variety of other parameters, such as the pulse wave velocity.

## Conflicts of Interest

The authors declare that they have no conflict of interest.

## References

1. Grimnes, S. Martinsen, O.G.: Bioelectricity and Bioimpedance Basics. 2nd edn. Academic Press: Cambridge, MA, USA, (2008).
2. Matthie, J.R.: Bioimpedance measurements of human body composition: critical analysis and outlook. Expert Rev Med Devices 5(2), 239-61 (2008).
3. Kusche, R., et al.: A FPGA-Based Broadband EIT System for Complex Bioimpedance Measurement - Design and Performance Estimation. Electronics 4(3), 507-25 (2015).

4. Northridge, D. B. et al.: Non-Invasive Determination of Cardiac Output by Doppler Echocardiography and Electrical Bioimpedance. British Heart Journal 63(2), 93–97 (1990).
5. Kaufmann, S., Malhotra, A., Ardelt, G., Ryschka, M.: A high accuracy broadband measurement system for time resolved complex bioimpedance measurements. Physiol Meas. 35(6), 1163-80 (2014).
6. Nyboer, J.: Electrical impedance plethysmography; a physical and physiologic approach to peripheral vascular study. Circulation 2(6), 811-21 (1950).
7. Kusche, R., Adornetto, T. D., Klimach, P., Ryschka, M.: A Bioimpedance Measurement System for Pulse Wave Analysis. In: 8th International Workshop on Impedance Spectroscopy, Chemnitz (2015).
8. Akay, M.: Wiley Encyclopedia of Biomedical Engineering. 1st edn. John Wiley & Sons Inc., Hoboken, New Jersey (2006).
9. Yazicioglu, R.F., van Hoof, C., Puers, R.: Biopotential Readout Circuits for Portable Acquisition Systems. 1st edn. Springer Netherlands (2009)
10. Park, J.H., Jang, D.G., Park, J. W., Youm, S.K.: Wearable Sensing of In-Ear Pressure for Heart Rate Monitoring with a Piezoelectric Sensor. Sensors 15(9), 23402–17 (2015).
11. Kusche, R., et al.: An in-ear pulse wave velocity measurement system using heart sounds as time reference. Current Directions in Biomedical Engineering, 1(1), 366-70 (2015).
12. Kusche, R., Hauschild, S., Ryschka, M.: Galvanically Decoupled Current Source Modules for Multi-Channel Bioimpedance Measurement Systems. Electronics 6, 90 (2017).